\documentclass[11pt]{article}
\usepackage{amsmath,amssymb,color,graphics,epsfig}
\usepackage{amsfonts}
\newcommand{\be}{\begin{equation}}
\newcommand{\ee}{\end{equation}}
\newcommand{\bea}{\setlength\arraycolsep{2pt} \begin{eqnarray}}
\newcommand{\eea}{\end{eqnarray}}
\newcommand{\nn}{\nonumber}
\newcommand{\mm}{\mathrm}
\newcommand{\mc}{\mathcal}

\def\fft#1#2{{\frac{#1}{#2}}}

\def\0{{\sst{(0)}}}
\def\1{{\sst{(1)}}}
\def\2{{\sst{(2)}}}
\def\3{{\sst{(3)}}}
\def\4{{\sst{(4)}}}
\def\5{{\sst{(5)}}}
\def\6{{\sst{(6)}}}
\def\7{{\sst{(7)}}}
\def\8{{\sst{(8)}}}
\def\sst#1{{\scriptscriptstyle #1}}

\begin{document}

\begin{flushright}
%\hfill{KIAS-P12028}
 %\hfill{
%\bf hep-th/yymmnnn}
\end{flushright}

\vspace{25pt}
\begin{center}
{\large {\bf Criticality of global monopole charges in diverse dimensions}}

\vspace{10pt}
 Hong-Ming Cui and Zhong-Ying Fan
%Xing-Hui Feng$^{2}$ and Hong L\"u$^{2\,,3}$

\vspace{10pt}
{ Department of Astrophysics, School of Physics and Electronic Engineering, \\
 Guangzhou University, Guangzhou 510006, China }\\
%\smallskip
%{\it $^{2}$Department of Physics and State Key Laboratory of Nuclear Physics and Technology,\\}
%{\it Peking University, No.5 Yiheyuan Rd, Beijing 100871, P.R. China\\}
%\smallskip
%{\it $^{3}$Collaborative Innovation Center of Quantum Matter, No.5 Yiheyuan Rd,\\}
%{\it  Beijing 100871, P. R. China\\}

\vspace{40pt}

\underline{ABSTRACT}
\end{center}

In this work, we construct charged AdS black holes with a global monopole charge in diverse dimensions and study the thermodynamics. We find a critical monopole charge below which the solution exhibits Van-der Waals like behaviors. In the context of holography, this could be intepreted using the boundary degrees of freedoms. As an example, we study the phase diagram in the four dimensions analytically. We further analyze the microstructures of the solution using Ruppeiner geometry. We find that repulsive interactions dominates for black holes in a wide range of temperatures. However, around the critical point, attractive interactions is dominant and the Ruppeiner scalar curvature shows universal behaviors: it has a critical exponent $2$ and coefficient $-1/8$ in diverse dimensions. Universality of the results is interpreted from the scaling behavior of free energy near the critical point for Van-der Waals like fluids.

%\vfill {\footnotesize  Email: fanzhy@pku.edu.cn\,.}

\thispagestyle{empty}

\pagebreak

\tableofcontents
\addtocontents{toc}{\protect\setcounter{tocdepth}{2}}

%%%%%%%%%%%%%%%%%%%%%%%%%%%%%%%%%%%%%%%%

%\newpage
%%%%%%%%%%%%%%%%%%%%%%%%%%%%%%%%%%%%%%%%

%\vspace{2cm}
\section{Introduction}
Einstein's gravity with a negative cosmological constant admits black hole solutions in asymptotically anti-de Sitter (AdS) spacetimes. Generally the solutions have richer phase structures relative to the asymptotically flat counterparts. In the past decade, people was interested in studying extended thermodynamics of AdS black holes, in which coupling constants in the Lagrangian were taken as thermodynamic variables. In \cite{Kubiznak:2012wp}, by treating the cosmological constant as a thermodynamic variable, referred to as thermodynamic pressure $P$, it was established that charged AdS black holes exhibit $P-V$ criticality, similar to the liquid-gas transition of Van-der Waals fluids. More interestingly, Ruppeiner geometry associated to thermodynamic fluctuations of the volume (or pressure) partly illustrates the microstructures of the black holes \cite{Wei:2015iwa,Wei:2019uqg}. 
Recently the thermodynamics of charged AdS black holes was further extended by introducing central charges in the boundary CFTs \cite{Cong:2021fnf,Cong:2021jgb}. This corresponds to varying both the cosmological constant and the Newton's gravitational constant. An advantage of this is the Van-der Waals like behaviors of AdS black holes can be interpreted in the context of holography. Generalization to higher curvature theories can be found in \cite{Cui:2024cnj,Kumar:2022afq}.

In this work, we would like to construct global monopoles swallowed by charged AdS black holes in diverse dimensions and study the thermodynamics. Previously, the topic was partly studied in \cite{Ahmed:2016ucs,Zheng:2024glr} in the four dimensions but using a ghost-like Goldstone field. This is however less interesting in physics since the solution is generally unstable. We resolve this issue by using a canonical multiplet and obtain the correct solution. Similar to the asymptotically flat case \cite{Barriola:1989hx}, the global monopole produces a strong gravitational field but has little contributions to the black hole mass. However, the metric becomes globally singular because of a deficit solid angle. We define the conical deficit (up to a constant factor) to be the {\it global monopole charge}. We find that the solution exhibits new Van-der Waals like behaviors associated to this new charge and its conjugate. The critical point depends on the ratio of the electric charge $Q$ to (certain power of) the central charge $C$ in the boundary. As a consequence, all dual fluids with a same ratio will have the same critical point, leading to a critical line on the $Q-C$ plane.   

The paper is organized as follows. In section 2, we consider Einstein-Maxwell theory with a cosmological constant, which is minimally coupled to a multiplet scalar field and construct charged AdS black holes with a global monopole charge in diverse dimensions. In section 3, we derive the first law for the solutions and extend the thermodynamics appropriately. In section 4, we study criticality of global monopole charges in the four dimensions in details. We derive the coexistence curve and analyze the phase diagram analytically. We further study the microstructures of the solution using the Ruppeiner geometry. In section 5, we generalize the results to higher dimensions and compute various critical exponents. We briefly conclude in section 6.

\section{Charged black holes with a global monopole}
Global monopoles form as a result of breaking of a global gauge symmetry. The simplest model producing a global monopole is described by a multiplet scalar field (\ref{multiplet}). To find charged AdS black hole solutions outside a global monopole, we consider Einstein-Maxwell theory with a cosmological constant, minimally coupled to a multiplet $\phi^a$ in diverse dimensions 
\bea
I=\int d^{D}x\,\sqrt{-g}\,\left( \fft{1}{16\pi G}(R-2\Lambda)-\fft{1}{16\pi}F^2+\mc{L}[\phi^a]\right) \,,
\eea
where $a=1\,,2\,,\cdots\,,D-1$, $\Lambda=-\fft{(D-1)(D-2)}{2L^2}$ is the cosmological constant and  $L$ is the AdS radius. The Lagrangian of the scalar field is given by 
\bea\label{multiplet}
\mc{L}=-\fft12 (\partial \phi^a)^2-V(\phi^a)\,,\quad V(\phi^a)=\fft{\lambda}{4}(\phi^a\phi^a-\eta^2)^2 \,,
\eea
where $\eta$ is the scale of gauge symmetry breaking. The model has a global O($D-1$) symmetry, which will be spontaneously broken to O($D-2$). For a typical unification scale $\eta\sim 10^{16}\mm{Gev}\ll \ell_P^{-1}$ , where $\ell_P$ is the Plank length. This is the four dimensional case. Generalization to diverse dimensions, the scale $\eta$ should be bounded by $\ell_P^{-\fft{D-2}{2}}$ on dimensional grounds. 

The Einstein field equations is given by
\bea
R_{\mu\nu}-\fft12R g_{\mu\nu}+\Lambda g_{\mu\nu}=8\pi G \left(T_{\mu\nu}^{\mm{EM}}+T_{\mu\nu}^{GM}\right) \,,
\eea
where
\bea
&&T_{\mu\nu}^{\mm{EM}}=\fft{1}{8\pi}\left( F_{\mu\alpha}F^{\alpha}_{\,\,\,\nu}-\fft14 g_{\mu\nu}F^2\right)\,,\nn\\
&&T_{\mu\nu}^{GM}=\partial_\mu \phi^a\partial_\nu\phi^a-g_{\mu\nu}\left[\fft12 (\partial \phi^a)^2+V(\phi^a) \right]\,.
\eea
 The equation of motion of the scalar field reads
\be \square\phi^a=\fft{\partial V}{\partial \phi^a}  \,.\ee
The generally static spherically symmetric charged solution takes the form of 
\bea
&& ds^2=-\tilde{f}(r)dt^2+f(r)^{-1}dr^2+r^2d\Omega_{D-2}^2 \,,\nn\\
&&A=a(r)\,dt\,,\qquad \phi^a=\eta h(r)\fft{x^a}{r}\,,
\eea
where $d\Omega_{D-2}^2$ is the metric of a unit $\mm{S}^{D-2}$ and  $x^ax^a=r^2$. Generally speaking, full solutions to $h(r)$ and the metric functions $\tilde{f}(r)\,,f(r)$ can only be found numerically. However, since most of the monopole energy is concentrated in a small region near the core, whose size is essentially determined by the scale of gauge symmetry breaking $\delta\sim \eta^{-\fft{2}{D-2}}$, we may take the monopole core $\delta\ll r_+$ in diverse dimensions, where $r_+$ denotes the black hole horizon radius. Under this approximation, $h(r)= 1$ and the potential $V(\phi^a)$ vanishes outside the horizon. The stress tensor of the multiplet greatly simplifies to
\bea
T^{0\mm{(GM)}}_{\,\,\,0}=T^{1\mm{(GM)}}_{\,\,\,1}=-\fft{(D-2)\eta^2}{2r^2}\,,\qquad T^{i\mm{(GM)}}_{\,\,\,i}=-\fft{(D-4)\eta^2}{2r^2} \,,
\eea
where $i=1\,,2\,,\cdots\,,D$. Consequently, the metric function $\tilde{f}(r)=f(r)$ so that the electrostatic potential can be solved immediately from Maxwell's equation
\be  a(r)=-\fft{4\pi Q}{(D-3)\Omega_{D-2}\, r^{D-3}}\,,\ee
where $\Omega_{D-2}=2\pi^{(D-1)/2}/\Gamma\left(\fft{D-1}{2}\right)$ is the volume of a unit $\mm{S}^{D-2}$ and $Q$ is the electric charge carried by the black hole. Finally, the metric function can be solved as
\be f(r)=1-2\zeta-\fft{16\pi GM}{(D-2)\Omega_{D-2}\, r^{D-3}}+\fft{32\pi^2 GQ^2}{(D-1)(D-3)\Omega_{D-2}^2\,r^{2D-6}}+\fft{r^2}{L^2}  \,,\label{fsol}\ee 
where
\bea
\zeta=\fft{4\pi G\eta^2}{D-3} \,,
\eea
is referred to as the {\it global monopole charge}. Clearly the presence of a global monopole charge does not contribute to the black hole mass\footnote{Precisely speaking, the monopole has energy $M_{\mm{core}}\sim \eta^{\fft{2}{D-2}}$ which is much less than the black hole mass.}. All its physical effect is inducing a conical deficit at asymptotical infinity. To see this, let $t\rightarrow t/\sqrt{1-2\zeta}$, $r\rightarrow r\sqrt{1-2\zeta}$, one finds asymptotically
\be  ds^2\Big|_{r\rightarrow \infty}=-\left( \fft{r^2}{L^2} +1\right)dt^2+\fft{dr^2}{ \fft{r^2}{L^2} +1}+(1-2\zeta)\,r^2d\Omega_{D-2}^2 \,.\ee
Clearly there exists a conical deficit for the solid angle $\Delta=\Omega_{D-2}\,\zeta$. Besides, the monopole charge should be upper bounded as $\zeta<1/2$, which translates into $\eta<\sqrt{\fft{D-3}{8\pi}}\,\ell_P^{-\fft{D-2}{2}}$.

\section{Extended black hole thermodynamics}
We proceed to study thermodynamics of the charged AdS black holes with a global monopole. Since the monopole charge is also an integration constant of the solution, the first law of thermodynamics acquires an extra contribution from a new pair of conjugates $(U\,,\zeta)$
\bea
dM=TdS+\Phi dQ+Ud\zeta\,,
\eea
where  $U$ is the chemical potential dual to $\zeta$.
The various quantities are given by
\bea
&&M=\fft{(D-2)\Omega_{D-2}\,r_+^{D-3}}{16\pi G}\left(Z+\fft{r_+^2}{L^2}\right)+\fft{2\pi Q^2}{(D-3)\Omega_{D-2}\,r_+^{D-3}}\,,\nn\\
&&T=\fft{1}{4\pi r_+}\left[(D-3)Z+(D-1)\fft{r_+^2}{L^2}\right]-\fft{8\pi GQ^2}{(D-2)\Omega_{D-2}^2r_+^{2D-5}}\,,\nn\\
&&S=\fft{\Omega_{D-2}\,r_+^{D-2}}{4G}\,,\quad \Phi=\fft{4\pi Q}{(D-3)\Omega_{D-2}\,r_+^{D-3}}\,,\quad U=-\fft{(D-2)\Omega_{D-2}}{8\pi Gr_+^{D-3}}\,,
\label{thermalquantity}\eea
where $Z\equiv 1-2\zeta$ is a useful combination here and after. It turns out that there exists a critical monopole charge below which the solution shows new Van-der Waals like behaviors in diverse dimensions.

However before studying this in details, we would like to extend the black hole thermodynamics by introducing the thermodynamic pressure \cite{Kubiznak:2012wp} and the central charge \cite{Cong:2021fnf} dual to the boundary
\bea
&&P=-\fft{\Lambda}{8\pi G}\,,\nn\\
&&C=\fft{kL^{D-2}}{16\pi G}\,,
\label{CP}\eea
where $k$ is a constant depending on details of holographic systems. 
The first law is extended to
\bea
dM=TdS+\Phi dQ+Ud\zeta+V_CdP+\mu dC\,,
\eea
where the thermodynamic volume $V_C$ and the chemical potential $\mu$ dual to the central charge are given by
\bea
&&V_C=\fft{\Omega_{D-2} r_+^{D-1}}{D(D-1)}+\fft{\Omega_{D-2}L^2Zr_+^{D-3}}{D(D-1)}+\fft{32\pi^2GL^2Q^2}{D(D-1)(D-2)\Omega_{D-2}r_+^{D-3}}\,,\nn\\
&&\mu=-\fft{2(D-1)\Omega_{D-2}r_+^{D-1}}{DkL^D}+\fft{2\Omega_{D-2}Zr_+^{D-3}}{DkL^{D-2}}+\fft{64\pi^2GQ^2}{D(D-2)\Omega_{D-2}kL^{D-2}r_+^{D-3}}\,.
\eea
Here the thermodynamic volume $V_C$ should not be confused with the original one defined at constant $G$ \cite{Cvetic:2010jb}. The two volumes are related by
\be V=V_C+\mu\fft{\partial C}{\partial P}\Big|_G=V_C-\fft{(D-2)\mu C}{2P} \,.\ee
Though it is widely believed that $V$ obeys the reverse isoperimetric inequality, this new thermodynamic volume $V_C$ generally does not \cite{Cong:2021fnf}. The Smarr relation is given by
\be M=\fft{D-2}{D-3}\big(TS+\mu C\big)+\Phi Q-\fft{2}{D-3}PV_C \,,\ee
where the monopole charge does not have any contributions (since $\zeta$ is dimensionless).

Dependence of the various thermodynamic quantities on the pressure $P$ and the central charge $C$ can be seen by using the inverse formula
\bea\label{inverse}
G=\fft{1}{16\pi}\left(\fft{k}{C}\right)^{2/D}\left[\fft{(D-1)(D-2)}{P}\right]^{1-2/D}\,,\quad
L=\left[\fft{(D-1)(D-2)C}{kP}\right]^{1/D}\,.
\eea
Since we are interested in $U-\zeta$ criticality, we will  work in canonical ensemble which has fixed charges $(Q\,,C\,,P\,,Z)$ or equivalently $(Q\,,G\,,L\,,\zeta)$. The equation of state of the dual fluid can be formally written as $T=T(U\,,\zeta)$. Explicit expressions for the equation of state will be presented later when we focus on certain dimensions. Here we shall point out that the existence of a critical point and the Van-der Waals-like behaviors associated to the monopole charge does not rely on the extended thermodynamics introduced above. All we have done in this part is trying to interpret the results in the context of holography, as will be shown later.

\begin{figure}
  \centering
  \includegraphics[width=300pt]{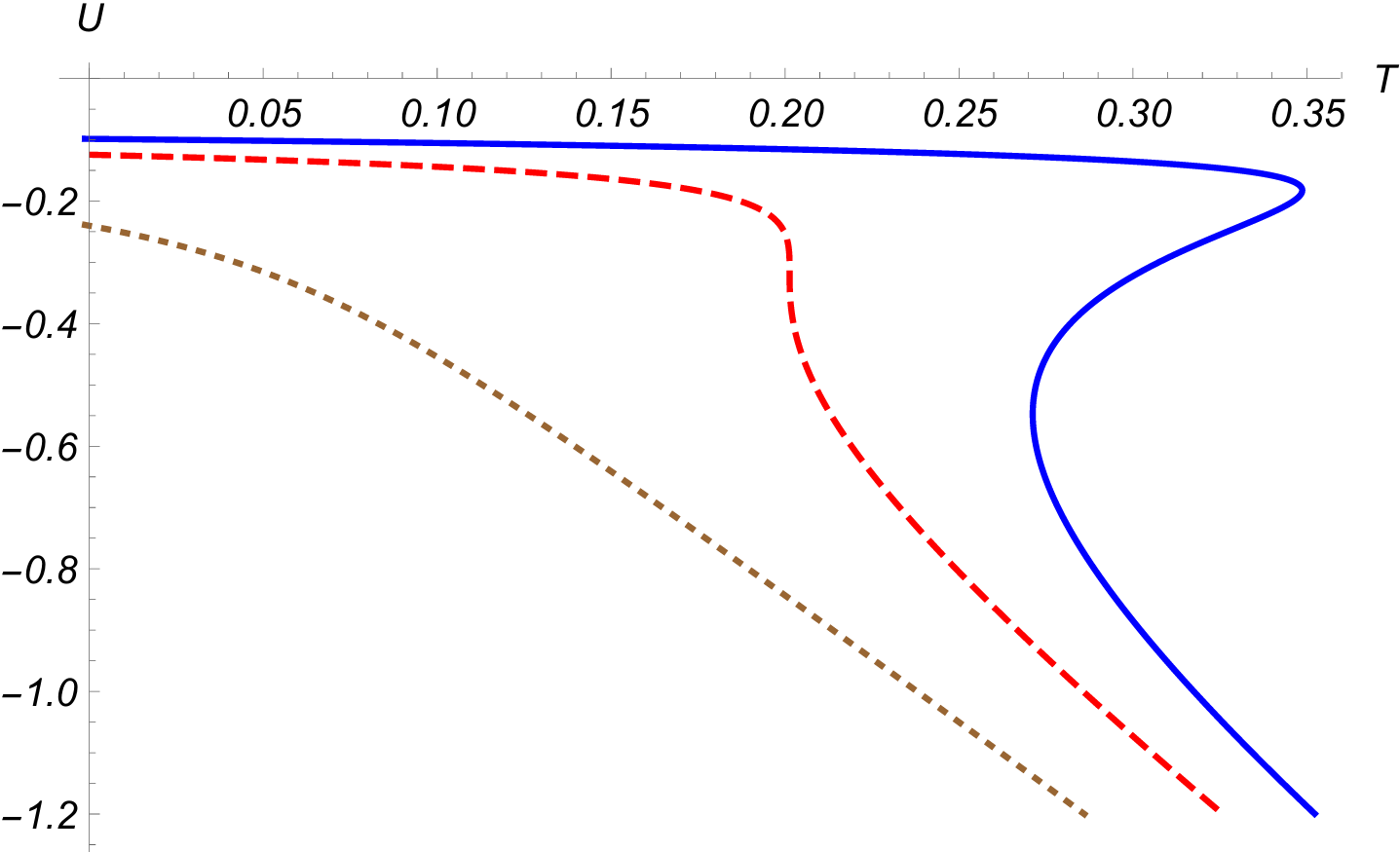}
  \caption{The chemical potential $U$ is displayed as a function of $T$ for fixed monopole charge. The various curves correspond to: $\zeta<\zeta_c$ (solid blue), $\zeta=\zeta_c$ (dashed red) and $\zeta>\zeta_c$ (dotted brown). Here we have set $G=1$, $L=1$ and $Q=0.1$.}
  \label{4D U-T}
\end{figure}

\section{Monopole charge criticality in 4-dimensions}
To study the $U-\zeta$ criticality in details, we consider the four dimensions at first.  The black hole solution reads
\bea
ds^2=-f(r)dt^2+\fft{1}{f(r)}dr^2+r^2d\Omega^2\,, \nn\\
f(r)=1-2\zeta-\fft{2GM}{r}+\fft{GQ^2}{r^2}+\fft{r^2}{L^2}\,.
\eea
The relevant thermodynamic quantities are given by
\bea
&&M=\fft{r_+^3}{2GL^2}+\fft{Z}{2G}r_{+}+\fft{Q}{2r_+}\,,\nn\\
&&T=\fft{3r_+}{4\pi L^2}+\fft{Z}{4\pi r_+}-\fft{GQ^2}{4\pi r_+^3}\,,\nn\\
&&S=\fft{\pi r_+^2}{G}\,,\quad U=-\fft{r_+}{G}\,.
\eea
Furthermore, the equation of state $T(U\,,\zeta)$ can be written explicitly as 
\be\label{eos4}
T=-\fft{2Z}{U}\sqrt{\fft{2PC}{3k}}-\fft{3kU}{64\pi^2 C }+\fft{32\pi P C Q^2}{3 k U^3}\,,
\ee
where we have adopted the relation $U=-8\pi r_+\sqrt{\fft{2PC}{3k}}$. Using inflection point condition $0=\partial T/\partial r_+=\partial^2 T/\partial r_+^2$, we obtain the critical point
\bea
&&r_c=\fft{3^{1/4}\sqrt{Q}}{2^{3/4}\pi^{1/4}P^{1/4}}\,,\nn\\
&&Z_c=\fft{3Q}{2}\sqrt{\fft{k}{\pi C}}\,,\nn\\
&&T_c=\fft{P^{1/4}}{6^{1/4}\pi^{5/4}}\sqrt{\fft{kQ}{C}}\,.
\eea
It is immediately seen that the critical monopole charge depends on the ratio of electric charge to (square root of) the central charge in the boundary. In fact, this gives a critical line on the $Q-C$ plane: all the dual fluids with a same ratio $Q/\sqrt{C}$ will have the same critical point. Interestingly, the product 
\be \fft{T_c r_c}{Z_c}=\fft{1}{3\pi}\,,\ee
is a constant, independent of the charges on the boundary (but unfortunately we do not find a proper interpretation for this). Define dimensionless variables $t=T/T_c$, $u=U/U_c$, $z=Z/Z_c$. The equation of state translates into the ``law of corresponding state"
\bea\label{eosu4}
t=\alpha u+\fft{\beta}{u}+\fft{\gamma}{u^3}\,,
\eea
where the various parameters are given by
\be \alpha=\fft{3}{8}\,,\quad \beta=\fft{3z}{4}\,,\quad \gamma=-\fft{1}{8} \,.\label{abcparameter}\ee
This is a very useful relation in solving the coexistence curve as well as computing various critical exponents.
\begin{figure}[h]
  \centering
  \includegraphics[width=250pt]{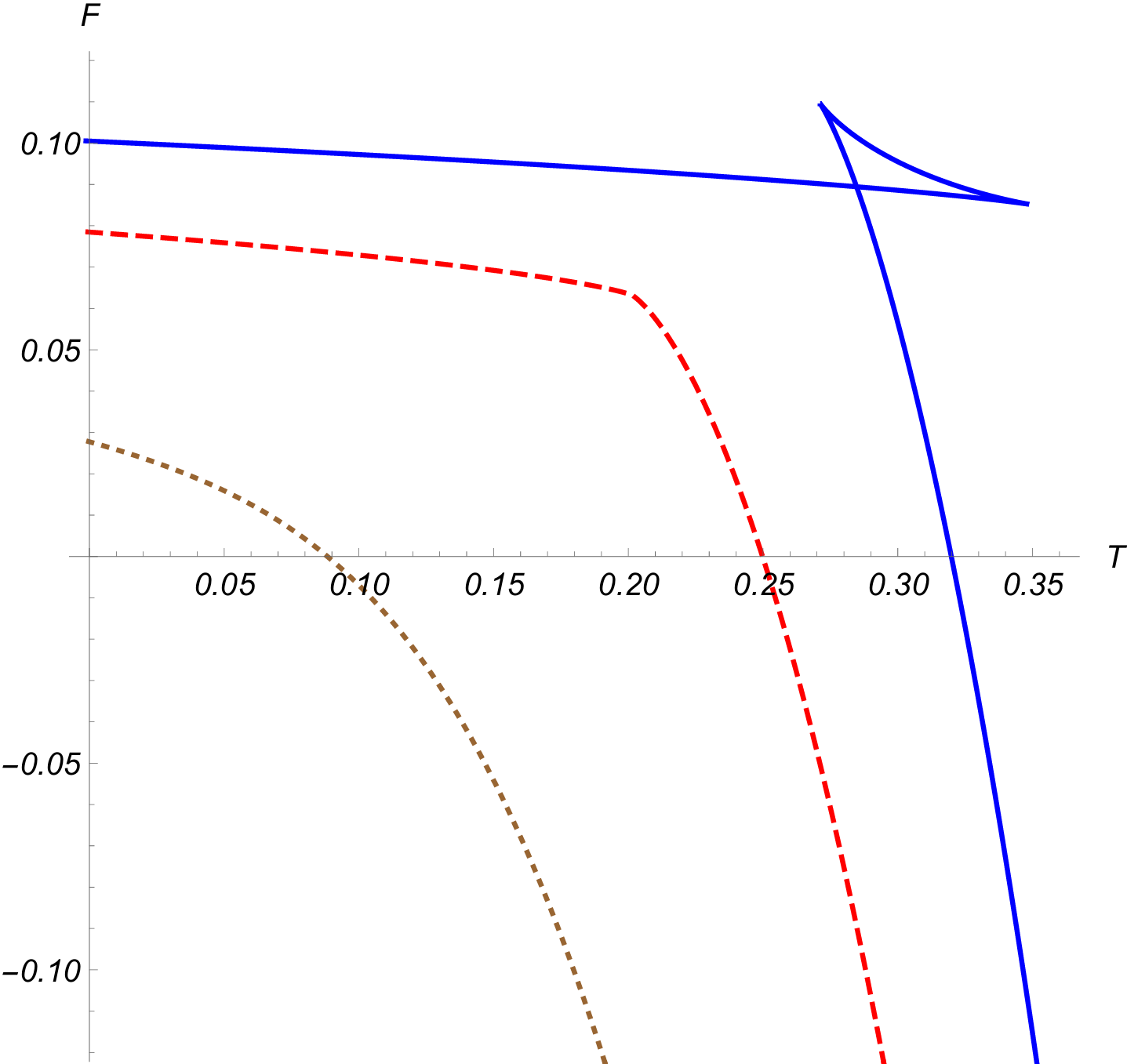}\,\,\,\,\,
  \caption{$F-T$ diagram is displayed as a function of $\zeta$ for fixed $G=1$, $L=1$ and $Q=0.1$. The various curves correspond to: $\zeta<\zeta_c$ (solid blue), $\zeta=\zeta_c$ (dashed red) and $\zeta>\zeta_c$ (dotted brown). We have set $k=16\pi$.
  }
  \label{4D FT}
\end{figure}

The $U-T$ diagram for a given monopole charge is depicted in Fig. \ref{4D U-T}. It follows that for $\zeta<\zeta_c$, the chemical potential $U$ has multi-values for a given temperature. This indicates a first order transition between a pair of black holes, across which $U$ changes discontinuously. To ensure this, we plot the $F-T$ diagram in Fig. \ref{4D FT}, where $F=M-TS$ is the free energy, given by
\be
F=-\fft{2}{3}\pi Pr_+^3+2\pi Z r_{+}\sqrt{\fft{2CP}{3k}}+\fft{3Q^2}{4r_+}\,.
\ee
It is easily seen that for a smaller monopole charge $\zeta<\zeta_c$, the swallowtail behavior emerges. It implies a first order phase transition for a small monopole charge but no such transition for a large monopole charge. The situation is somehow similar to the electric charge criticality \cite{wald2}, in which the transition between multi-phases is only possible when the number of charged carriers is sufficiently small, though microstructures of the two cases are quite different.

\subsection{Coexistence curve and the phase diagram}

A remarkable feature of the four dimensional solution is the coexistence curve can be derived analytically. We will generalize the method developed in \cite{wald4} and adopt the Maxwell's area law on  the $T-S$ plane.
\begin{figure}
  \centering
  \includegraphics[width=300pt]{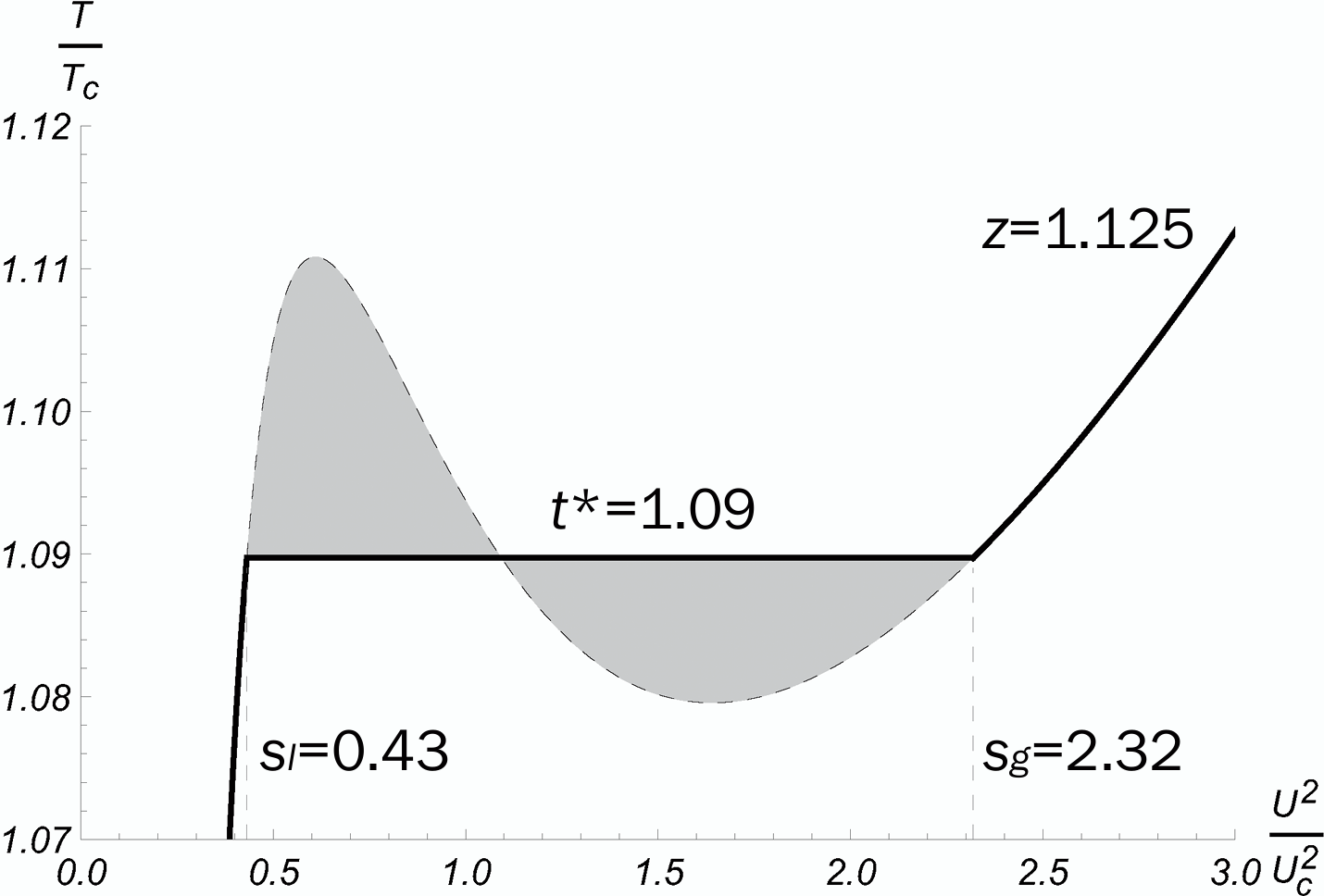}
  \caption{Maxwell's area law on the $T-S$ graph. The oscillatory region is unphysical and should be replaced by an isotherm $t^*$, according to the area law. The rectangular region below the isotherm is the coexistence region. We have set $z=\fft{9}{8}$, that is $\zeta=\fft{18\zeta_c-1}{16}$ and $t^*\approx1.09$.}
  \label{T-S AL}
\end{figure}
 For this purpose, it will be convenient for us to work with the law of corresponding state (\ref{eosu4}). However, to keep the discussions below as general as possible, we will not specify the parameters $(\alpha\,,\beta\,,\gamma)$ until the end of this subsection. The equal area law gives
\bea
t^*\left(u_g^2-u_l^2\right)=\int^{u^2_g}_{u^2_l}t(u)du^2\,,
\eea
where $u_l$ and $u_g$ ($u_l<u_g$) denotes the normalized chemical potential of the small and the large black hole on the the isotherm $t=t^*$, respectively (see Fig. \ref{T-S AL} for an illustration). Integrating the law of corresponding state yields
\be
t^*=\fft{2}{u_g+u_l}\left[\fft{\alpha}{3}(u^2_g+u^2_l+u_g\,u_l)+\beta+\fft{\gamma}{u_g\,u_l}\right]\,.
\label{t*1}
\ee
On the other hand, the two phases must have the same temperature
\bea
&&t^*=t(u_g)=t(u_l)\,,\nn\\
&&t^*=\fft{t(u_g)+t(u_l)}{2}\,,
\eea
which leads to
\bea
&&\left(u_g+u_l\right)^2-\fft{\alpha}{\gamma}(u_g u_l)^3+\fft{\beta}{\gamma}(u_g u_l)^2-u_g u_l=0\,,\nn\\
&&t^*=\fft{u_g+u_l}{2}\left[\alpha+\fft{\beta}{u_g\,u_l}+\gamma\fft{(u_g+u_l)^2-u_g\,u_l}{(u_g\,u_l)^3}\right]\,.
\label{t*2}
\eea
It follows that the equations (\ref{t*1}) and (\ref{t*2}) can be solved analytically by taking $u_gu_l$ and $u_g+u_l$ as independent variables. One finds
\bea
\left[\alpha (u_g u_l)^2+3\gamma\right]\left[\beta u_g\,u_l-\alpha (u_g u_l)^2+3\gamma\right]=0 \,,
\eea
which gives two set solutions
\bea
\alpha (u_g u_l)^2+3\gamma=0\quad&\Rightarrow&\quad  u_gu_l=\sqrt{-\fft{3\gamma}{\alpha}}\,,\label{phyc}\\
 \beta u_g\,u_l-\alpha (u_g u_l)^2+3\gamma=0 \quad&\Rightarrow& \quad u_g u_l=\fft{\beta\pm \sqrt{\beta^2+12\alpha\gamma}}{2\alpha}\,\,.
\label{unphyc}\eea
However, the second solution is unphysical since it will lead to $u^2_g=u^2_l=\fft{\beta\pm\sqrt{\beta^2+12\alpha\gamma}}{2\alpha}$. From (\ref{phyc}), we obtain the physical solutions
\bea
u^2_{g,\,l}=\fft{1}{2\alpha}\left[3\beta-4\sqrt{-3\alpha\gamma}\pm\sqrt{3(3\beta-2\sqrt{-3\alpha\gamma})(\beta-2\sqrt{-3\alpha\gamma})}\,\right]\,,\label{coesgl}\eea
where the sign ``$+$'' (``$-$'') corresponds to the saturated large (small) black hole, respectively.
Substituing the parameters (\ref{abcparameter}) into (\ref{coesgl}) and (\ref{t*1}), we finally obtain 
\bea
u_{g,l}=\sqrt{3z-2\pm\sqrt{(3z-1)(3z-3)} }\,,
\label{sgl}\eea
and the coexistence curve on the $t-z$ plane
\bea
t^*(z)=\sqrt{\fft{3z-1}{2}}\,.\label{coetz}
\eea
Notice that $z> 1/3$ is guaranteed by the upper bound of the monopole charge $\zeta<1/2$.
\begin{figure}
  \centering
  \includegraphics[width=300pt]{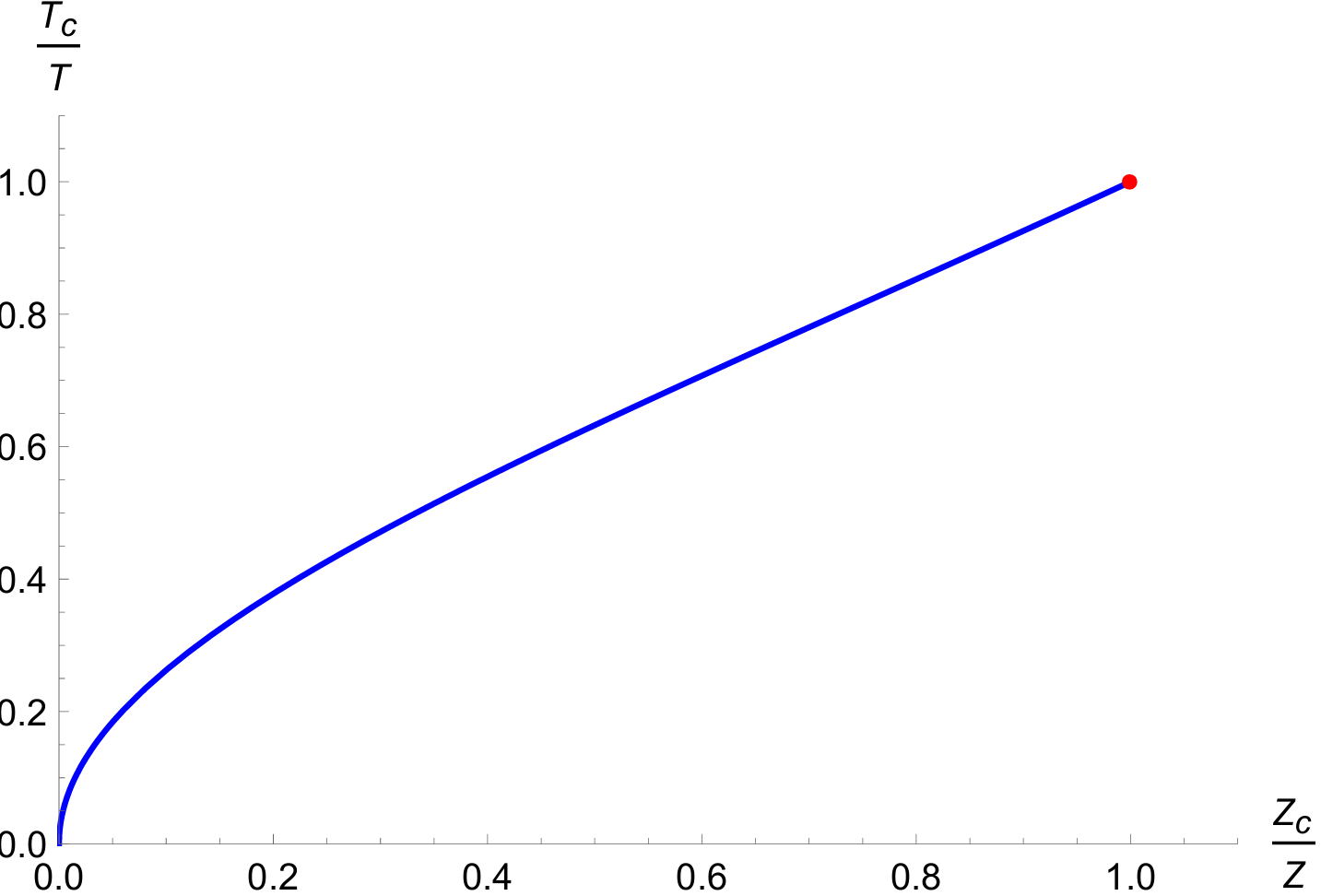}
  \caption{The coexistence curve of for charged AdS black holes with a global monople on the $t-z$ plane.}
  \label{ts(z)}
\end{figure}
\begin{figure}
  \centering
  \includegraphics[width=300pt]{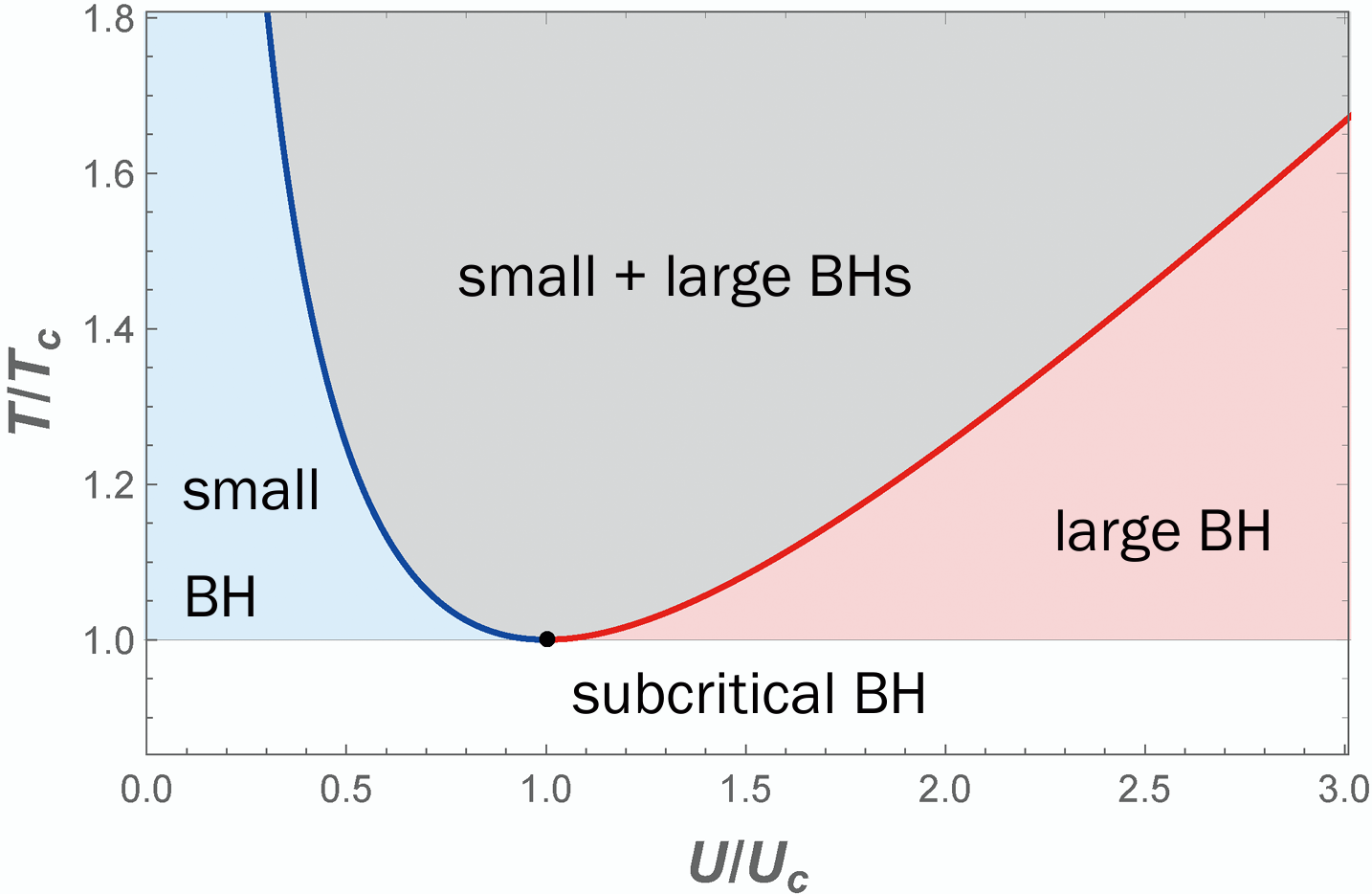}
  \caption{Phase diagram of charged AdS black holes with a global monopole on the $t-u$ plane. Blue and red solid curves are separated by the critical point (black dot), corresponding to the saturated small and large black holes respectively. The three phases are represented by the shaded regions with different color.}
  \label{Z pd}
\end{figure}
The coexistence line is depicted in Fig. \ref{ts(z)}. Notice that the transition occurs for $T>T_c$, which is equivalent to $\zeta<\zeta_c$. Furthermore, combing (\ref{sgl}) and (\ref{coetz}), we deduce the coexistence curve on the $t-u$ plane 
\be t^*(u)=\fft{u^2+1}{2u} \,.\ee
With these results in hand, we are ready to draw the phase diagram, as shown in Fig. \ref{Z pd}. It is easily seen that given a small monopole charge $\zeta<\zeta_c$, there exists three phases for the solution:\nn\\
$\bullet$ for $u<u_l$, there is a single small black hole in the ``liquid-phase";\nn\\
$\bullet$ for $u>u_g$, there is a single large black hole in the ``gas-phase";\nn\\
$\bullet$ for $u_l<u<u_g$, the black hole is in a coexisting ``liquid-gas phases", transiting from one phase to another.\nn\\
These results are standard for Van-der Waals-like fluids. 

\subsection{Rupepeiner geometry and microstructures}
\begin{figure}[h]
  \centering
  \includegraphics[width=300pt]{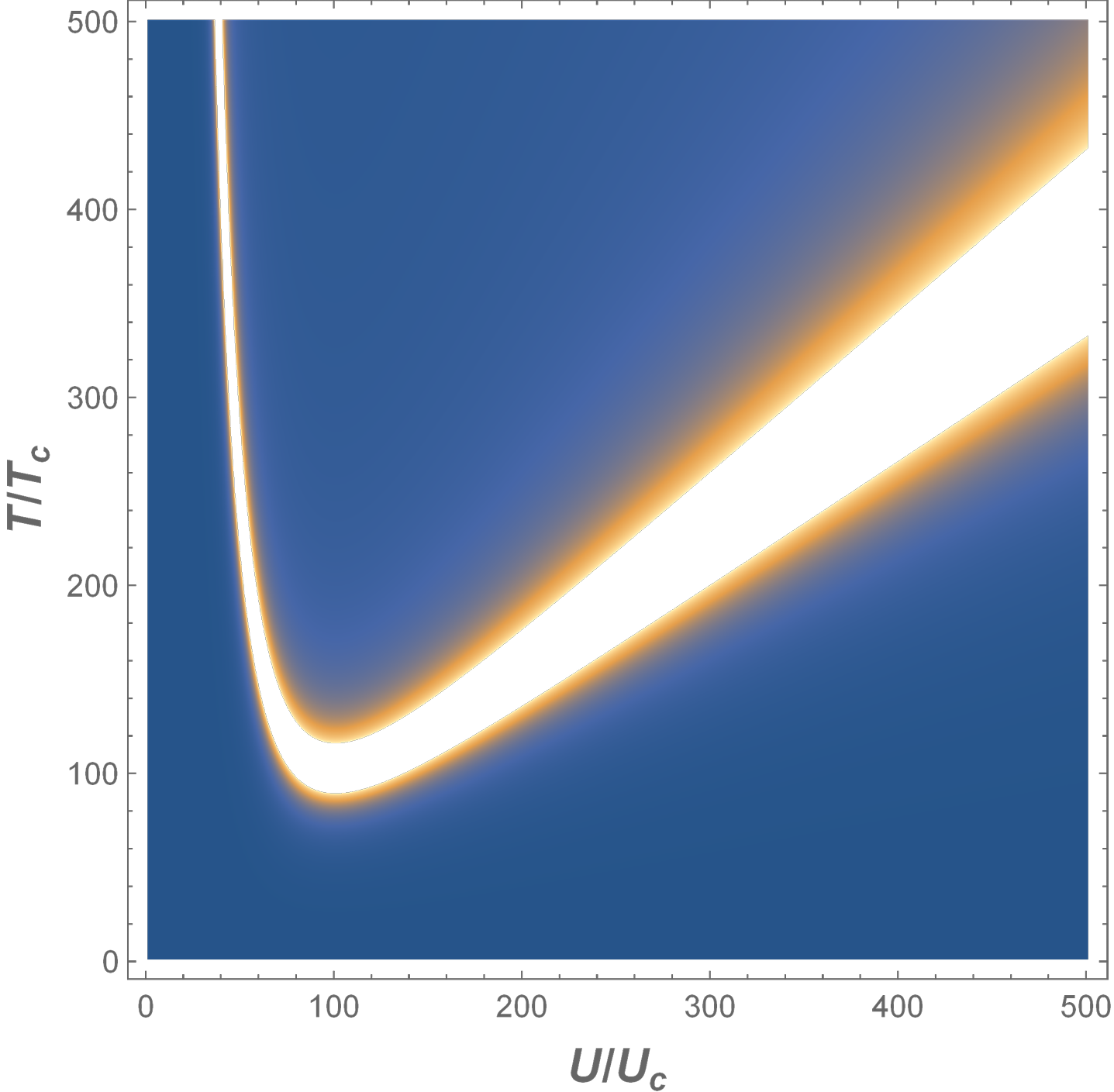}
  \includegraphics[width=32pt]{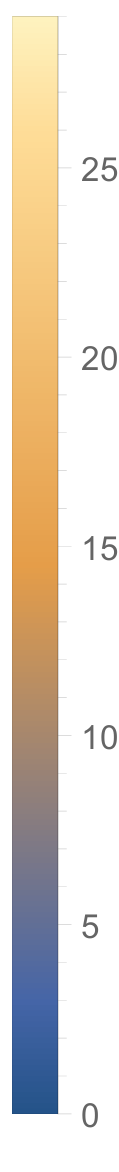}
  \caption{Density plot for the normalized scalar curvature $R_N$ in the $D=4$ dimensions.}
  \label{Z geo}
\end{figure}
Having established the phase diagram, we move to study microstructures of the dual fluids using Ruppeiner geometry. In thermodynamic fluctuation theory, the distance between two neighboring fluctuation states $\Delta l^2$ is related to the probability of finding the macroscopic system between the neighboring states $P(x^0\,,\,\cdots,x^N)\varpropto e^{-\fft{1}{2}\Delta l^2}$, where
\bea
\Delta l^2=-\fft{\partial^2S}{\partial x^{\mu}\partial x^{\nu}}\Delta x^{\mu}\Delta x^{\nu}\,.
\eea
We choose the temperature $T$ and the chemical potential $U$ to be the independent fluctuation variables. The line element reads
\bea\label{dl}
dl^2=\fft{C_{U}}{T^2}dT^2-\fft{1}{T}\left(\fft{\partial \zeta}{\partial U}\right)_{T}dU^2\,,
\eea
where $C_U=T\left(\fft{\partial S}{\partial T}\right)_U$ is the heat capacity at constant $U$. However, since $U\propto\sqrt{S}$, the heat capacity $C_U$ vanishes identically. To solve this issue, we adopt the trick in \cite{Wei:2019uqg}: taking $C_U$ to be a nonzero constant at first and then sending $C_U\rightarrow 0$. In this limit, the Ruppeiner scalar curvature diverges as $R\propto 1/C_U$. Then we define a normalized scalar curvature $R_N\equiv R C_U$ and propose that it describes the microstructures of the dual fluids: a positive (negative) $R_N$ indicating that repulsive (attractive) interactions dominates.

Evaluation of the normalized scalar curvature yields
\bea
R_N=-\fft{(3u^4+1)(8tu^3-3u^4-1)}{2(4tu^3-3u^4-1)^2}\,.
\eea
The behavior of $R_N$ is depicted in Fig. \ref{Z geo}. Roughly speaking, $R_N$ is very small in most of the parameters space. However, close to the curve
\bea
t_{div}=\fft{3u^4+1}{4u^3}\,,
\eea
the curvature $R_N$ changes dramatically. This curve exactly corresponds to a divergent heat capacity $C_\zeta$ defined at constant monopole charge, signaling critical point of the transition. On this curve, $R_N$ goes to negative infinity. Besides, there exists a sign-changing curve across which $R_N$ changes the sign 
\bea
t_0=\fft{3u^4+1}{8u^3}\,.
\eea
It follows that for $t>t_0$, $R_N<0$, implying the attractive interactions dominates in this region. Otherwise, $R_N$ will be positive and the repulsive interactions dominates instead. 
\begin{figure}[h]
  \centering
  \includegraphics[width=300pt]{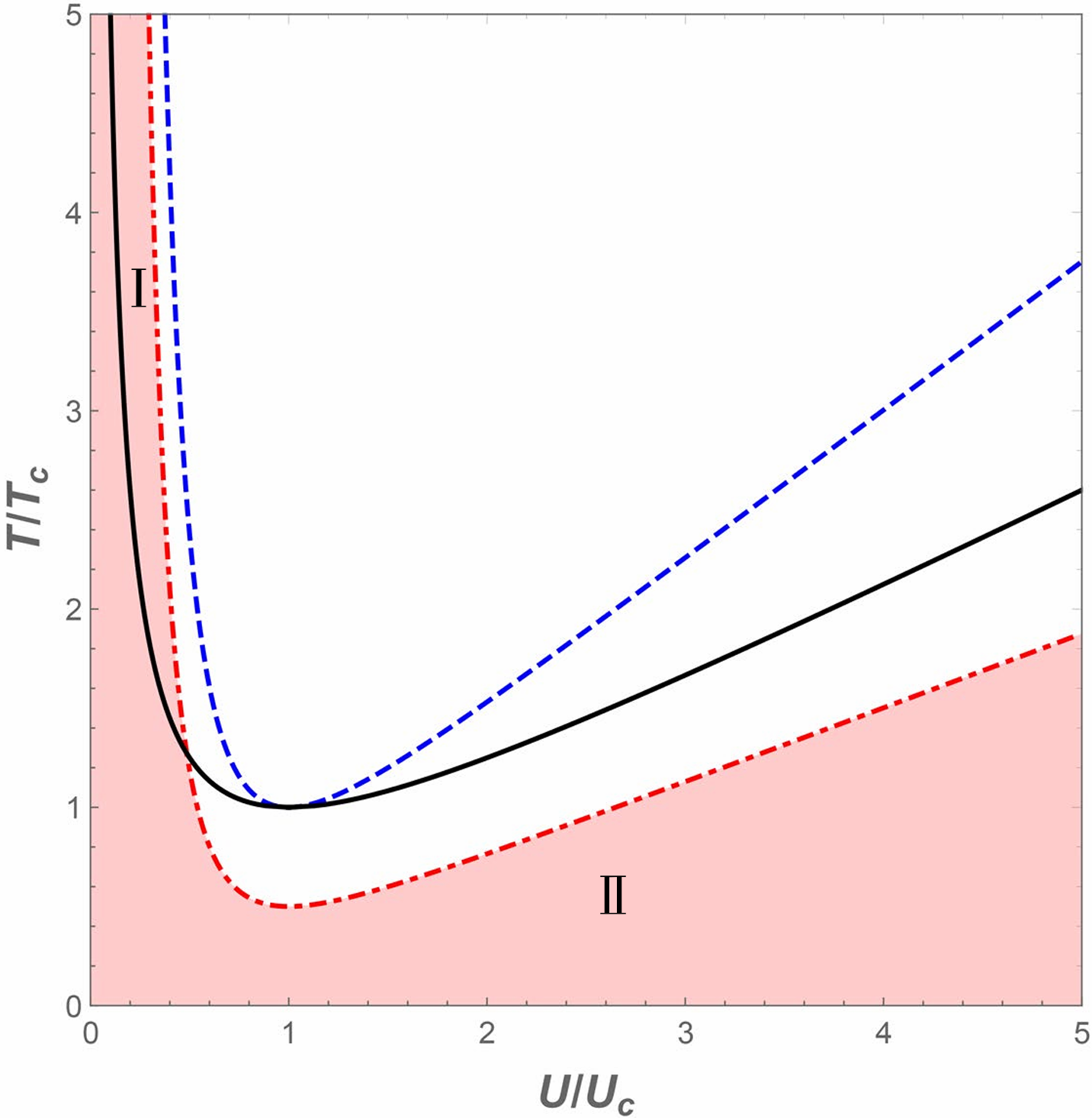}
  \caption{Characteristic curves for  $D=4$ dimensions. The black solid line is the coexistence curve, and the red dot-dashed line is sign-changing curve corresponding to $t_0$, on which $R_N=0$. The blue dashed line corresponds to the temperature $t_{div}$, on which $R_N\rightarrow-\infty$. In the shaded region $R_N>0$ and otherwise $R_N<0$.}
  \label{4D RN analyze}
\end{figure}
In Fig. \ref{4D RN analyze}, we illustrate the coexistence curve (black solid line), the sign-changing curve (red dot-dashed line) as well as the diverging curve (blue dashed line), respectively. The shaded (unshaded) regions correspond to a positive (negative) curvature $R_N$, indicating that the repulsive (attractive) interactions dominates. However, any features above the coexistence curve are tentative since the equation of state we adopted is inapplicable in the coexistence region. The results that we have confidence to state are those below the coexistence curve. In region II, repulsive interactions dominates in the microstructures, including most of the saturated small black holes. However, around the critical point, attractive interactions is dominant. To explain this, let us expand the scalar curvature in powers of $\delta t=t-1$. We deduce
\bea R_N=-\fft{1}{8\delta t^2}\pm \fft{1}{(2\delta t)^{3/2}}+\cdots \,,\eea
where ``+/-'' sign corresponds to the saturated small/large black holes, respectively. To leading order $R_N$ has a critical exponent $2$ and the coefficient 
\bea
\lim_{\delta t\rightarrow 0}R_N\delta t^2=-\fft{1}{8} \,.\label{R4crit}
\eea 
In fact, these results are universal to charged AdS black holes with a global monopole in diverse dimensions and match with the Van-der Waals fluid perfectly \cite{Wei:2019uqg}. Universality of the result could be interpreted from the scaling behavior of free energy near the critical point, see the next section.
\begin{figure}
  \centering
  \includegraphics[width=200pt]{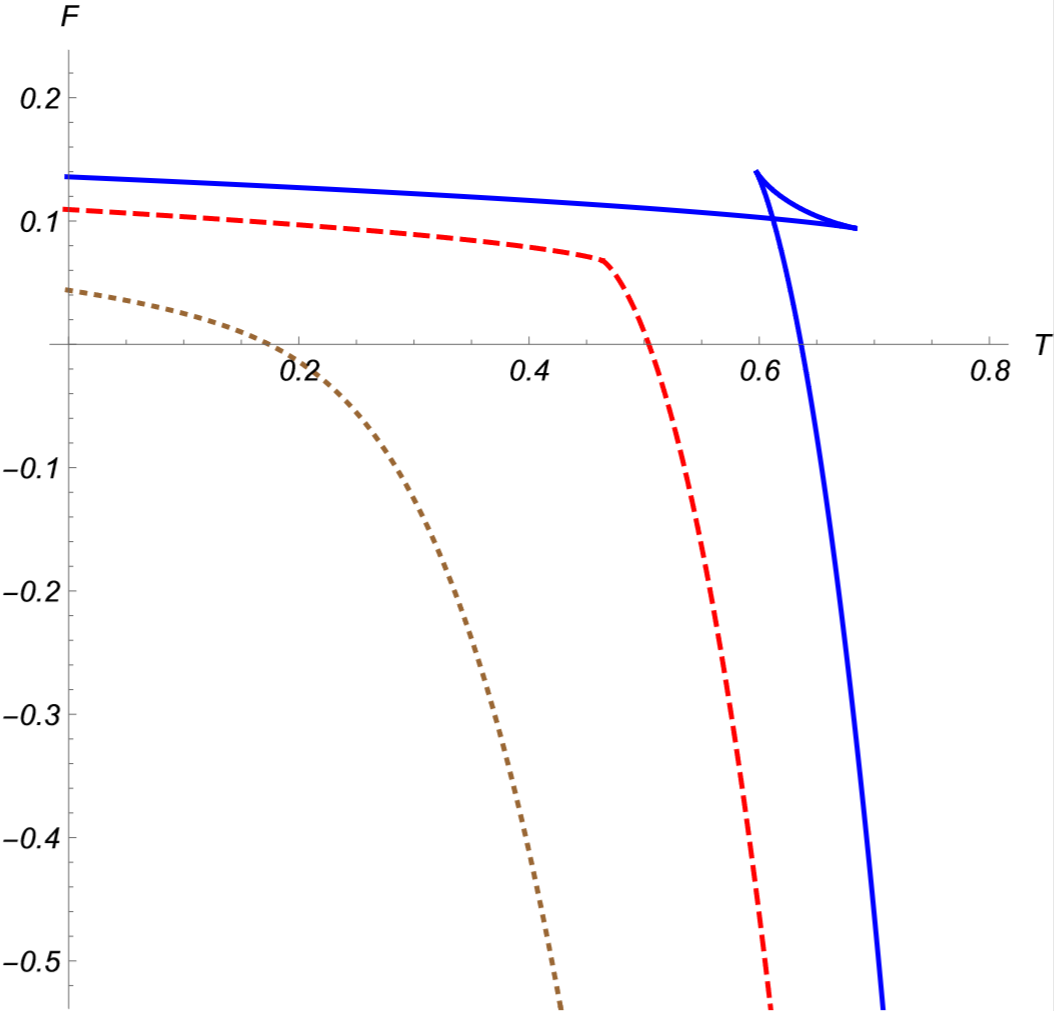}
  \includegraphics[width=200pt]{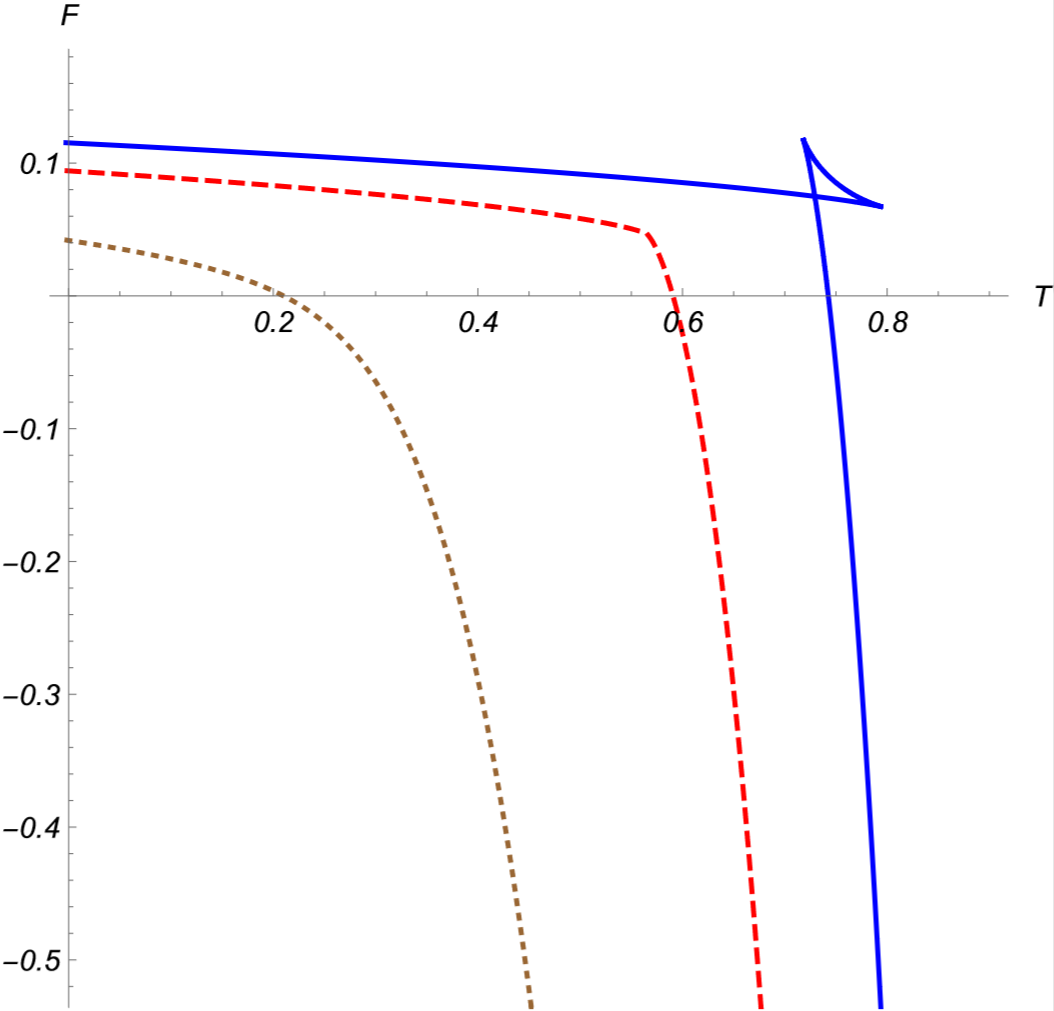}
  \caption{ The F-T digram for $D=5$ (left) and $D=6$ (right). In both panels, $z>1$ (solid blue), $z=1$ (dashed red) and $z<1$ (dotted brown).}
  \label{56F-T}
\end{figure}

\section{Monopole charge criticality in higher dimensions}
We continue studying the critical phenomenon of monopole charges in higher dimensions. The equation of state $T=T(U\,,\zeta)$ is listed as follows
\bea T=\fft{1}{4\pi r_+}\left[(D-3)Z+(D-1)\fft{r_+^2}{L^2}\right]-\fft{8\pi GQ^2}{(D-2)\Omega_{D-2}^2r_+^{2D-5}}\,,\eea
where 
\bea r_+^{D-3}=-\fft{(D-2)\Omega_{D-2}}{8\pi G U} \,.\eea
Evaluation of the critical point using the inflection point condition yields
\bea
r_c&=&\left[\fft{2(D-3)(2D-5)\pi Q^2}{\Omega_{D-2}^2P}\right]^{\fft{1}{2D-4}}\,,\\
Z_c&=&\fft{2\big[(D-2)(D-1)\big]^{\fft{D-2}{D}}[2\pi(D-3)(2D-5)]^{\fft{1}{D-2}}}{(2D-5)\pi}\fft{k^{\fft{2}{D}}Q^{\fft{2}{D-2}}}{\Omega_{D-2}^{\fft{1}{D-2}}C^{\fft{2}{D}}P^{\fft{D-4}{D(D-2)}}}\,,\\
T_c&=&\fft{\big[(D-2)(D-1)\big]^{\fft{D-2}{D}}[2\pi(D-3)(2D-5)]^{\fft{1}{2D-4}}}{(2D-5)\pi}\fft{k^{\fft{2}{D}}Q^{\fft{1}{D-2}}}{\Omega_{D-2}^{\fft{1}{D-2}}C^{\fft{2}{D}}P^{\fft{3D-8}{2D(D-2)}}}\,.
\eea
Again the critical monopole charge depends on the ratio of the electric charge to (certain power of) the central charge: $Q/C^{\fft{D-2}{D}}$. In addition, the product
\bea
\fft{T_c r_c}{Z_c}=\fft{(D-3)^2}{(2D-5)\pi}\,,
\eea
turns out to be a constant in any given dimensions. 

Working with dimensionless variables  $t=T/T_c\,,u=U/U_c\,,z=Z/Z_c$, we deduce the law of corresponding state
\bea
t=\fft{2D-5}{4(D-2)}u^{\fft{1}{D-3}}+\fft{2D-5}{4(D-3)}zu^{-\fft{1}{D-3}}-\fft{u^{-\fft{2D-5}{D-3}}}{4(D-2)(D-3)}\,,
\eea
and the normalized free energy $f=F/F_c$ 
\bea f=\fft{D-1}{4}z\, u+\fft{D-1}{4(D-2)u}-\fft{(D-3)^2}{4(D-2)}\,u^{\fft{D-1}{D-3}} \,.\eea
The $F-T$ digrams for higher dimensional solutions are depicted in Fig. \ref{56F-T}. The existence of multi-phases and the transitions between them (the coexistence curve) for a small monopole charge $\zeta<\zeta_c$ as well as the full phase digram are quite similar to the four dimensional case. We shall not discuss them anymore. 

\begin{figure}
  \centering
  \includegraphics[width=195pt]{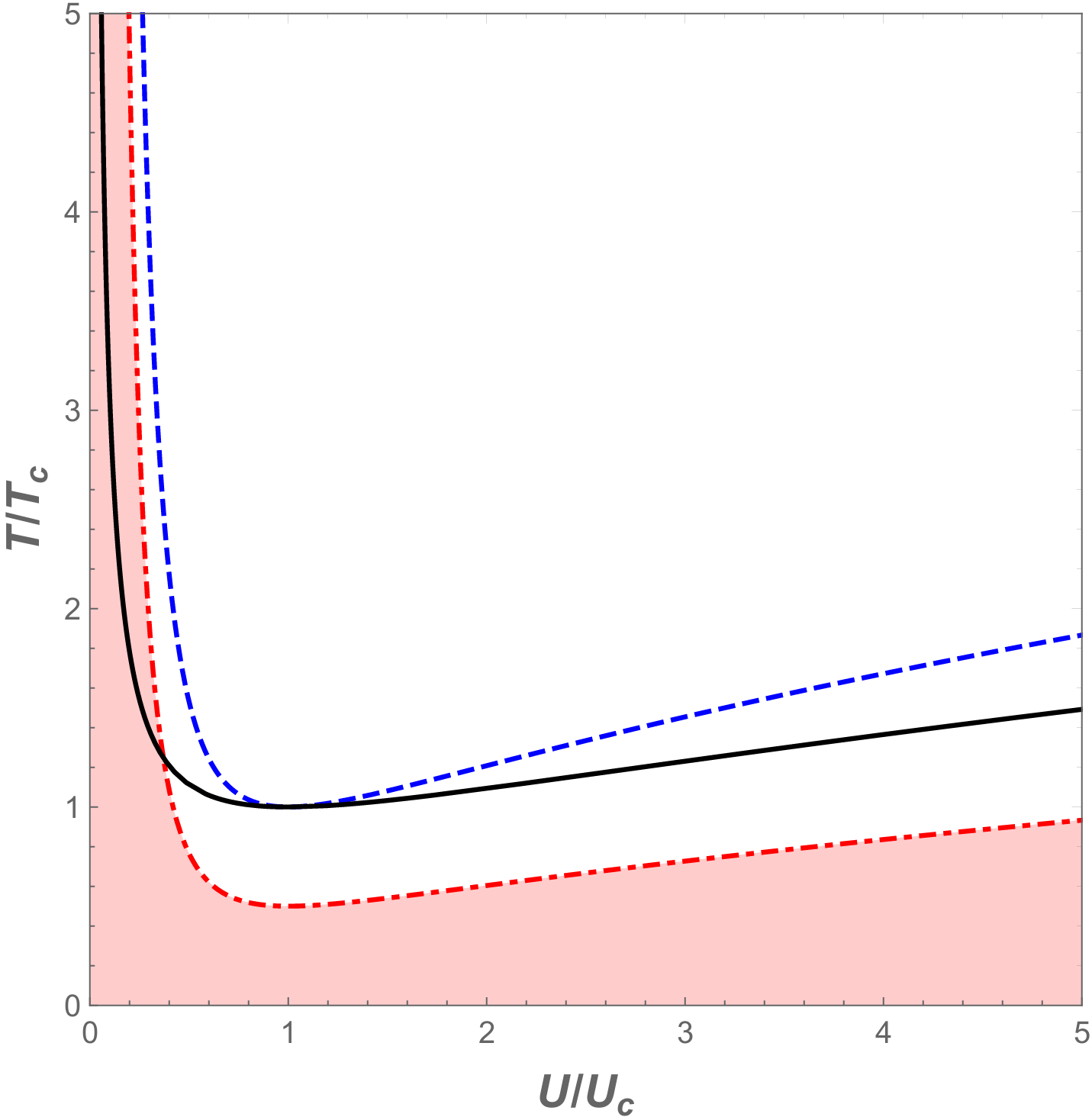}
  \includegraphics[width=195pt]{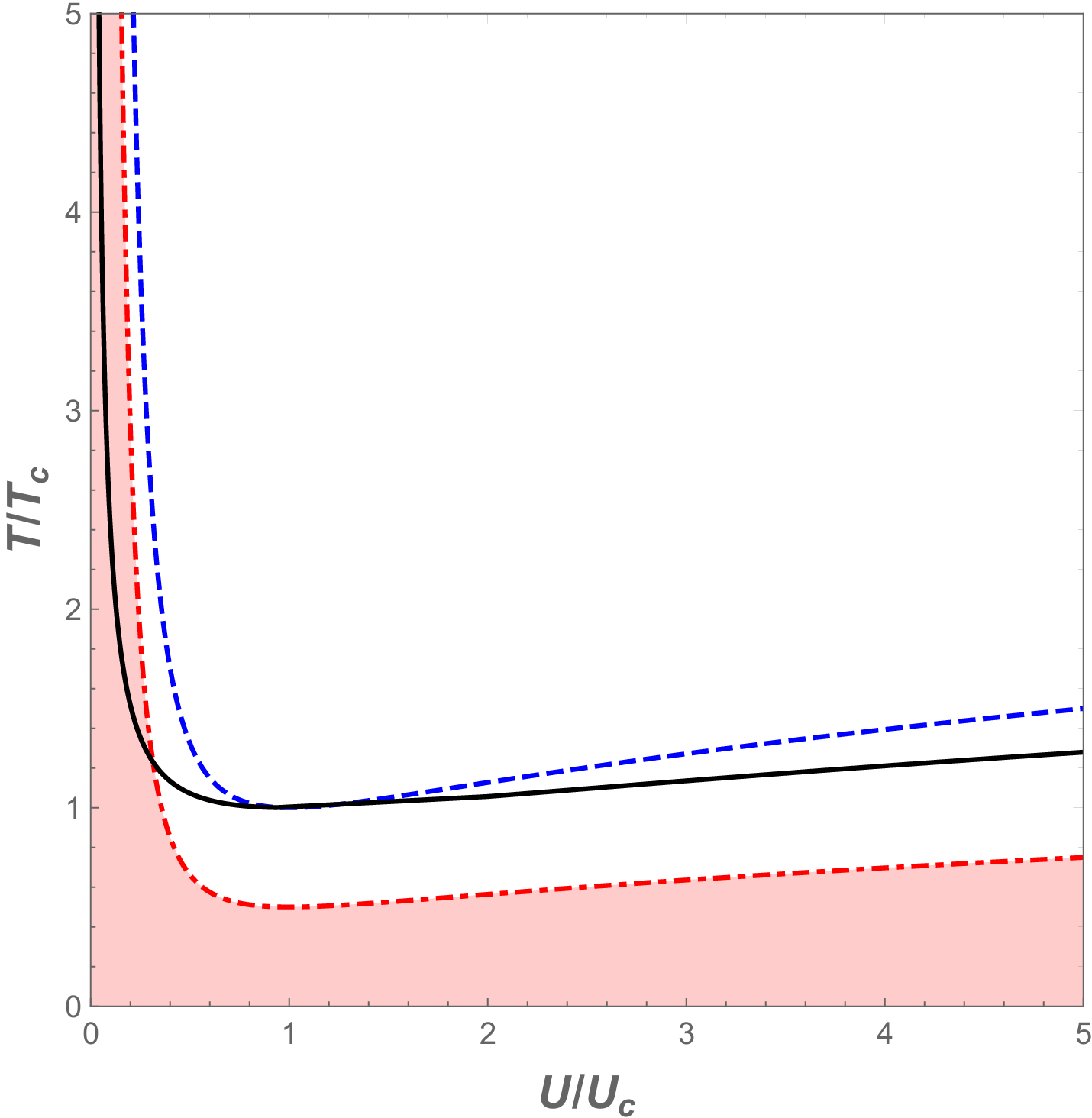}
  \caption{ Behavior of the normalized scalar curvature $R_N$ of $D=5$ (left) and $D=6$ (right) dimensions. All characteristic curves have the same meaning as those in Fig. \ref{4D RN analyze}. In the shaded region $R_N>0$ and otherwise $R_N<0$.}
  \label{56RN}
\end{figure}

To proceed, we study the microstructures for higher dimensional solutions using the Ruppeiner geometry. We deduce 
\bea R_N=-\fft{\left[(2D-5)\,u^{\fft{2(D-2)}{D-3}}+1 \right]\left[4(D-2)t\, u^{\fft{2D-5}{D-3}}-(2D-5)\,u^{\fft{2(D-2)}{D-3}}-1 \right]}{2\left[2(D-2)t\, u^{\fft{2D-5}{D-3}}-(2D-5)\,u^{\fft{2(D-2)}{D-3}}-1\right]^2} \,.\eea
Again there exists a diverging curve
\bea 
t_{div}=\fft{(2D-5)\,u^{\fft{2(D-2)}{D-3}}+1}{2(D-2)\, u^{\fft{2D-5}{D-3}}} \,,\eea
on which $R_N$ goes to negative infinity and a sign-changing curve
\bea t_0= \fft{(2D-5)\,u^{\fft{2(D-2)}{D-3}}+1}{4(D-2)\, u^{\fft{2D-5}{D-3}}} \,,\eea
across which the attractive and the repulsive interactions changes from one another. In Fig. \ref{56RN}, we depict these curves as well as the coexistence lines (which are computed numerically) for $D=5$ and $D=6$ dimensions, respectively. Similar to the four dimensional case, small black holes are dominated by repulsive interactions whereas the large black holes (at higher temperatures $t>t_0$) are dominated by attractive interactions. In particular, attractions always dominates around the critical point. The leading order behavior of the scalar curvature (\ref{R4crit}) near the critical point is universal to diverse dimensions.    

Universality of the result could be interpreted using the two-scale factor universality hypothesis \cite{stauffer}. Near the critical point, the singular part of the free energy connected to critical phenomena can be promoted to be a generalized homogeneous function \cite{widom,stanley}:
\bea F_s=c_1 |\delta t|^{\beta(\delta+1)} Y\big( c_2\, v |\delta t|^{-\beta \delta}\big)  \,,\eea
where $\beta\,,\delta$ are the critical exponents defined as usual (see next subsection), $v=u-1$ is the order parameter and $Y()$ is a smooth function, obeying the conditions $Y(0)<0$ and $Y'(0)=0$. Note that the constants $c_1\,,c_2$ are two scale factors, which are material dependent and differentiate between systems even in the same universality class. Actually, in modern approach to critical phenomena, the behavior of free energy can be elegantly explained using renormalization group analysis in quantum field theory.

By straightforward calculations, we deduce
\bea
&& R=\fft{\beta(\beta-\alpha)T_c}{2(2-\alpha)(1-\alpha)c_1 Y(0)}\,\delta t^{\alpha-2}\,,\\
&& C_U=-\fft{(2-\alpha)(1-\alpha)c_1 Y(0)}{T_c}\,\delta t^{-\alpha}\,,
\eea
where we have made use of the scaling equation $2-\alpha=\beta(\delta+1)$. It is immediately seen that the critical behaviors of both $R$ and $C_U$ depend on details of the fluids, through the scale factor $c_1$. However, the leading order behavior of the normalized curvature $R_N$ is universal 
\bea \lim_{\delta t\rightarrow 0}R_N \delta t^2=-\fft{\beta(\beta-\alpha)}{2} \,,\label{leadinggene} \eea
depending only on the critical exponents $\alpha\,,\beta$. For Van-der Waals like fluids, $\alpha=0\,,\beta=1/2$, this reproduces (\ref{R4crit}) exactly. The result is robust since neither a constant $C_U$ nor the limit $C_U\rightarrow 0$ plays an essential role.

\subsection{Critical exponents}
We move to compute the various critical exponents for $U-\zeta$ criticality in diverse dimensions. The exponents $\alpha$, $\beta$, $\gamma$, $\delta$ are defined as follows: \\
$\bullet$ Exponent $\alpha$ describes the behavior of the specific heat at constant $U$:
\bea
C_U=T\left(\fft{\partial S}{\partial T}\right)_U \propto \lvert\delta t\rvert^{-\alpha}\,.
\eea
$\bullet$ Exponent $\beta$ describes the behavior of the order parameter $u_g-u_l$ on the given isotherm
\bea
u_g-u_l \propto \lvert\delta t\rvert^{\beta}\,.
\eea
$\bullet$ Exponent $\gamma$ determines the behavior of the isothermal compressibility $\kappa_T$
\bea
\kappa_T=-\fft{1}{U}\left(\fft{\partial U}{\partial \zeta}\right)_T \propto \lvert\delta t\rvert^{-\gamma}\,.
\eea
$\bullet$ Exponent $\delta$ describes the following behavior on the critical isotherm $T=T_c$:
\bea
\lvert \zeta-\zeta_c \rvert \propto \lvert U-U_c \rvert^{\delta}\,.
\eea
Firstly, since the entropy $S$ is a function of $U$, the specific heat  $C_U$ vanishes identically. This implies $\alpha=0$.

To compute $\beta$, we first rewrite the law of corresponding state as 
\bea
z=\fft{4(D-3)}{2D-5}(\delta t+1)(\upsilon+1)^{\fft{1}{D-3}}-\fft{D-3}{D-2}(\upsilon+1)^{\fft{2}{D-3}}+\fft{1}{(D-2)(2D-5)(\upsilon+1)^2}\,.
\label{z(t,u)}\eea
Expanding the equation to relevant orders in the vicinity of the critical point yields
\bea
z=1+\fft{4D-12}{2D-5}\delta t+\fft{4}{2D-5}\delta t\upsilon-\fft{2}{3(D-3)^2}\upsilon^3+O(\delta t\upsilon^2\,,\,\upsilon^4)\,.
\label{expand z}\eea
During the phase transition, the temperature remains a  constant, which gives
\bea
z&&=1+\fft{4D-12}{2D-5}\delta t+\fft{4}{2D-5}\delta t\upsilon_g-\fft{2}{3(D-3)^2}\upsilon_g^3 \nn\\
&&=1+\fft{4D-12}{2D-5}\delta t+\fft{4}{2D-5}\delta t\upsilon_l-\fft{2}{3(D-3)^2}\upsilon_l^3\,,
\eea
Besides, using Maxwell's area law, one has 
\bea
0=\int^{\upsilon_g}_{\upsilon_l}\upsilon dz=\int^{\upsilon_g}_{\upsilon_l}\left[\fft{4\delta t}{2D-5}-\fft{2\upsilon^2}{3(D-3)^2}\right]\upsilon d\upsilon=0\,.
\eea
Combining the equations, one finds the solution is $\upsilon_g=-\upsilon_l\propto\sqrt{-\delta t}$. This leads to $\beta=\fft{1}{2}$. 

Differentiating (\ref{expand z}), we get
\bea
\left(\fft{\partial U}{\partial \zeta}\right)_T=\fft{2D-5}{4}\fft{\upsilon_c\,z}{z_c\,\delta t}\qquad\Rightarrow \qquad \gamma=1\,.
\eea

Finally, according to (\ref{expand z}), the critical isotherm $\delta t=0$ adheres to 
\bea
z-1=-\fft{2\upsilon^3}{3(D-3)^2}\qquad\Rightarrow \qquad\delta=3\,.
\eea
This completes our derivations.

\section{Conclusion}
In this work, we constructed charged AdS black holes with a global monopole in diverse dimensions. The presence of a monopole charge $\zeta$ and its thermodynamic conjugate $U$ contributes to the first law of thermodynamics. We found that there exists a critical monopole charge $\zeta_c$ below which the solution exhibits Van-der Waals like behaviors. In the context of holography, this could be interpreted using the boundary degrees of freedoms. The critical point depends on the ratio of the electric charge to (certain power of) the central charge.

As an example, we derived the coexistence curve in the four dimensions and studied the phase diagram analytically. The results of higher dimensions are qualitatively similar. We also investigated the microstructures of the solution using thermodynamic geometry. We defined a normalized scalar curve $R_N$ and established that the leading order behavior (\ref{R4crit}) near the critical point is universal to diverse dimensions. Universality of the results is interpreted from the scaling behavior of free energy near the critical point for Van-der Waals like fluids. All these results improve our understanding of the critical phenomenon for charged AdS black holes with a global monopole charge significantly.

\section*{Acknowledgments}
Z.Y. Fan was supported in part by the National Natural Science Foundations of China with Grant No. 11805041 and No. 11873025 and also supported in part by Guangzhou Science and Technology Project 2023A03J0016.

\end{document}